\documentclass{llncs}

\usepackage[T1]{fontenc}
\usepackage{graphicx}
\usepackage{booktabs}
\usepackage{array}
\usepackage{amsmath}
\usepackage{url}
\usepackage{placeins}
\usepackage{xcolor}

\newcommand{\auc}{AUC}
\newcommand{\xlmrbase}{\mbox{XLM-R text baseline}}

\begin{document}

\title{Why Alzheimer's Speech Screening Fails to Generalize: Bridging the Deploym1ent Gap via Cross-Corpus Evidence Anchoring}
\titlerunning{Why Alzheimer's Speech Screening Fails to Generalize}
\author{Zijian Lu\inst{1} \and Sizhe Liu\inst{3} \and Yin Zhang\inst{3} \and Jixuan Deng\inst{3} \and Xinrong Lin\inst{2} \and Xinchen Yuan\inst{3} \and Chicheng Jin\inst{3} \and Yiping Zuo\inst{1} \and Yuanchao Li\inst{4}\thanks{Corresponding author}}
\authorrunning{Z. Lu et al.}
\institute{Nanjing University of Posts and Telecommunications, Nanjing, China\\
\and Shouyi Technology, Hefei, China\\
\and University of Science and Technology of China, Hefei, China\\
\and University of Cambridge, Cambridge, United Kingdom\\
\email{18818732360@163.com, yl809@cam.ac.uk}}

\maketitle

\begin{abstract}
Speech-based screening is a promising, non-invasive approach for detecting Alzheimer's disease and related cognitive risks. However, models trained on a single domain often generalize poorly to unseen languages, tasks, or recording protocols. This paper investigates this deployment gap using a leave-one-corpus-out evaluation across four distinct datasets. Among 70 interpretable speech and language features, 59 exhibit direction conflicts between healthy control and cognitive risk groups across corpora, with pause, silence, and speech rate showing high protocol sensitivity. Furthermore, while the \xlmrbase\ achieves strong average performance, its Area Under the ROC Curve (AUC) drops to 0.520 on the weakest held-out domain. A standard GroupDRO baseline reaches a 0.766 mean speaker AUC and a 0.504 worst-domain AUC under the same protocol. To address this, we propose a fusion method that integrates \xlmrbase\ scores with evidence anchors selected during training. Balanced fusion achieves a 0.785 mean speaker AUC, while anchor-heavy fusion raises the worst-case speaker AUC to 0.615. This work highlights the need to audit feature transferability and report worst-case domain robustness in cognitive speech screening.

\end{abstract}

\section{Introduction}

Speech offers a low-cost, non-invasive window into cognitive decline. Language production relies on memory, attention, lexical access, discourse planning, and executive control --- functions that are frequently impaired in Alzheimer's disease and mild cognitive impairment \cite{roark2011spoken,fraser2016linguistic}. Unlike neuroimaging or extended clinical batteries, spontaneous speech can be collected remotely, tracked longitudinally, and analyzed with both interpretable markers and machine learning models.

Two paradigms dominate the field. The first designs interpretable features such as pauses, speech rate, silence ratio, lexical diversity, and discourse cohesion. ASR-derived measures have been used to detect mild cognitive impairment from spontaneous speech \cite{roark2011spoken,toth2018asr,liu2025beyond}, temporal and acoustic parameters offer clinically meaningful insight into dementia speech \cite{meilan2014speech}, and narrative-speech linguistic features distinguish Alzheimer's disease from healthy controls in picture-description tasks \cite{fraser2016linguistic}, with lexical and discourse markers further predicting probable or future Alzheimer's disease status \cite{orimaye2017predicting,eyigoz2020linguistic}. The second paradigm uses deep representations such as BERT-style encoders \cite{devlin2019bert}, \xlmrbase\ \cite{conneau2020xlmr}, and self-supervised speech models like wav2vec 2.0 \cite{baevski2020wav2vec}, which capture complex patterns that escape single clinical markers \cite{li2025semi}.

While prior work confirms that speech carries a viable screening signal, it remains unclear whether this signal stays reliable once a screening system leaves a familiar domain. A deployed system meets new picture prompts, tasks, languages, microphones, segmentation policies, and ASR error profiles, and models tuned under random splits within one corpus risk fitting a mixture of disease evidence, task artifacts, and cohort bias. We show that both handcrafted markers and deep neural models fail to generalize reliably, though in different ways. Handcrafted features can be informative in one corpus yet heavily protocol-dependent in another: pause, silence, and speech rate are often treated as universal proxies for impairment, yet their direction can invert with task design, segmentation, or language, so single-marker rules are unsafe under domain shift. Deep representations pose a complementary risk, since strong average performance can mask severe failure on specific unseen corpora, which matters in practice because users experience their own deployment setting rather than a benchmark average. We therefore treat worst-case domain robustness as a primary evaluation criterion.

To address this, we propose a deployment-oriented framework for Alzheimer's disease speech screening under external corpus shift. The core idea is to use heterogeneous existing datasets to simulate future, unseen deployment environments during model development. In a Leave-One-Corpus-Out (LOCO) setup, the model learns from multiple source domains and is evaluated on one fully held-out corpus, mimicking transfer across new languages, tasks, or recording protocols. Our framework has three stages: it audits the cross-corpus reliability of speech and language markers, selects a compact set of robust evidence anchors from source data, and fuses these anchor scores with the \xlmrbase\ score. This shifts the goal from maximizing performance on familiar benchmarks toward ensuring that both the evidence and the predictions remain reliable in unseen deployment scenarios. Our contributions are threefold. First, an audit of 70 interpretable markers across four corpora shows that 59 exhibit direction conflicts between healthy control and cognitive risk groups, and these conflicts are structured, with lexical and POS markers stable while silence, speech rate, and pause duration are highly protocol-sensitive. Second, we show that deep representations do not fully mitigate this distribution shift, as the \xlmrbase\ achieves strong average performance but degrades sharply on the hardest held-out domain. Third, we introduce a transfer protocol that fuses audited evidence anchors with the \xlmrbase, where a balanced configuration yields the best mean \auc\ and an anchor-heavy configuration substantially raises performance on the weakest domain.

\section{Related Work}

\subsection{Speech and Language Markers for Cognitive Screening}

DementiaBank and the wider TalkBank repository provide the field's core transcribed clinical speech resources, mostly drawn from picture-description tasks \cite{macwhinney2011aphasiabank,lanzi2023dementiabank,talkbankpitt}. The Pitt Cookie corpus is a widely benchmarked English resource, while Mandarin corpora such as Chou offer important non-English validation \cite{talkbankchou,chou2024screening}. Reviews show that acoustic, prosodic, lexical, syntactic, and discourse markers all carry diagnostic value, but also report substantial variance across dataset characteristics, preprocessing, and evaluation protocols \cite{defuente2020review,martinez2021tenyears,li2026adreview}. Disfluency and interactional cues have further been used for spontaneous speech analysis, though these remain sensitive to transcription source and task format \cite{nasreen2021disfluency}.

Interpretable markers are clinically attractive because features like pause ratios or pronoun frequencies can be visualized and clinically inspected \cite{fraser2016linguistic}. \textbf{The critical risk is that clinical interpretability does not guarantee cross-corpus generalization.} Pause-based markers depend heavily on voice-activity-detection thresholds, microphone hardware, task, speaking time, and segmentation policy \cite{meilan2014speech}, and lexical markers are similarly sensitive to transcription pipeline, tokenization, and prompt design. We therefore separate two questions: whether a marker is statistically significant within a corpus, and whether its direction reliably transfers to an unseen domain.

\subsection{Shared Tasks and Cross-Corpus Evaluation}

Shared tasks such as ADReSS and ADReSSo standardized the field through balanced splits and unified baselines \cite{luz2020adress,luz2021adresso}. This was extended to multilingual mild cognitive impairment detection by TAUKADIAL \cite{barrera2024taukadial}, while NCMMSC 2021 introduced a Chinese Alzheimer's disease recognition evaluation with long and short speech tracks, serving as a rigorous test for Mandarin screening \cite{ncmmsc2021ad,chen2023raw}. Each corpus nonetheless remains tied to its own collection protocol, so robust evaluation should target held-out environments rather than in-distribution averages \cite{koh2021wilds}. Since a deployed system inevitably meets unfamiliar tasks and cohorts, we adopt a rigorous LOCO framework rather than random splitting.

\subsection{Deep Representations}

Pre-trained text encoders such as BERT \cite{devlin2019bert} and XLM-R \cite{conneau2020xlmr} offer strong linguistic modeling, with XLM-R particularly advantageous in cross-lingual settings, and have also been compared directly against traditional features for Alzheimer's disease detection \cite{balagopalan2020bert}. On the acoustic side, wav2vec 2.0 \cite{baevski2020wav2vec} and wavLM \cite{chen2022wavlm} provide effective self-supervised speech representations, and multimodal ADReSSo systems show that acoustic, lexical, disfluency, and pause features yield complementary gains \cite{rohanian2021acoustic}, echoing NCMMSC results where raw-waveform and ASR-driven deep architectures achieve competitive results on Chinese data \cite{chen2023raw,qin2021asr}. Despite this success, deep models face a severe deployment risk, since they easily overfit to corpus identity, transcription artifacts, channel conditions, or prompt structure \cite{balagopalan2020bert}, and self-supervised models also exhibit inconsistent behaviour across emotionally and cognitively distinct corpora \cite{li2023slt}. Rather than relying on the deep model alone, we treat XLM-R as an audited text stream and counteract its risk by fusing it with robust, cross-corpus evidence anchors.

\section{Data and Evaluation Protocol}

\subsection{Corpora}

We use four heterogeneous corpora spanning Mandarin and English, picture-description tasks, and connected clinical speech:

\begin{itemize}
    \item \textbf{NCMMSC:} A Mandarin dataset from the official Alzheimer's disease recognition evaluation challenge \cite{ncmmsc2021ad}. As it contains picture description, fluency test, and free conversation, we therefore label it a ``mixed'' task corpus.
    \item \textbf{Pitt Cookie:} From the DementiaBank Pitt Corpus, using spontaneous speech elicited via the ``Cookie Theft'' picture-description task \cite{talkbankpitt}.
    \item \textbf{TAUKADIAL Mandarin:} A subset of the multilingual mild cognitive impairment challenge, capturing connected speech from semi-structured clinical assessments \cite{barrera2024taukadial}.
    \item \textbf{Mandarin Chou:} A DementiaBank Mandarin corpus tracking language profiles of healthy controls and participants with amnestic mild cognitive impairment \cite{talkbankchou,chou2024screening}.
\end{itemize}

Speaker IDs are prefixed by corpus before evaluation, and for speakers with multiple samples, sample risk scores are averaged before computing speaker AUC. We standardize the task into a binary setting, healthy control (HC) versus cognitive risk, collapsing Alzheimer's disease, MCI, and other cognitive-impairment labels into a single risk class. Table~\ref{tab:data} summarizes the sample and speaker distributions.

\begin{table}[!ht]
\centering
\caption{Corpora used for leave-one-corpus-out evaluation. HC and risk are sample-level counts.}
\label{tab:data}
\footnotesize
\setlength{\tabcolsep}{4pt}
\begin{tabular*}{\textwidth}{@{\extracolsep{\fill}}lllrrrr@{}}
\toprule
Corpus & Language & Task & Samples & Speakers & HC & Risk \\
\midrule
NCMMSC & Mandarin & Mix & 246 & 116 & 90 & 156 \\
Pitt Cookie & English & Picture & 547 & 291 & 241 & 306 \\
TAUKADIAL Mandarin & Mandarin & Connected & 507 & 169 & 222 & 285 \\
Mandarin Chou & Mandarin & Picture & 261 & 87 & 120 & 141 \\
\midrule
Total &  &  & 1561 & 663 & 673 & 888 \\
\bottomrule
\end{tabular*}
\end{table}

\subsection{Features and Models}

The interpretable feature pool comprises 70 high-level speech and language features spanning temporal disfluency, pause dynamics, speech rate, acoustic intensity, lexical diversity, POS distributions, lexical concentration, repetition markers, information structure, and discourse cohesion, and a compact subset of 24 canonical features forms the traditional baseline marker set.

For deep baselines we use frozen pre-trained encoders with a lightweight downstream classifier: 768-dimensional \xlmrbase\ embeddings extracted from transcripts, and 1024-dimensional representations from the intermediate (12th) and final (24th) layers of a pre-trained wav2vec 2.0 model. No encoder is fine-tuned, so the comparison isolates domain robustness from representation drift.

All continuous features are preprocessed with parameters fitted strictly on the training partition, including winsorization at the 1st and 99th percentiles, median imputation, and $z$-score normalization. The downstream classifier is a weighted $L_2$-regularized logistic regression ($\lambda=1.0$), optimized via L-BFGS-B, with sample weights combining inverse class frequency and inverse per-speaker sample count. Decision thresholds are calibrated exclusively on training speakers, while all reported \auc\ scores remain threshold-free.

\subsection{Leave-One-Corpus-Out Protocol}

Our core evaluation is a Leave-One-Corpus-Out (LOCO) cross-validation, where each fold holds out one corpus as an unseen external evaluation set and aggregates the remaining three as the training partition. All pipeline components, including feature selection, scaling, model fitting, and threshold calibration, are computed exclusively on the training corpora to prevent target-domain leakage. The primary metric is speaker-level \auc, and we report both the macro-averaged mean speaker \auc\ across held-out folds and the worst-case domain speaker \auc.

As a standard domain-generalization comparator, we apply Group Distributionally Robust Optimization (GroupDRO) \cite{sagawa2020groupdro} to the same frozen XLM-R embeddings, treating each source corpus as a group. Group weights are updated by exponentiated gradient ascent on source-corpus logistic losses for 100 iterations ($\eta=0.1$), while retaining the same source-only preprocessing, sample weighting, and $L_2$ regularization ($\lambda=1.0$) as the \xlmrbase. This configuration is fixed across folds, and the held-out corpus is never used for group updates, model fitting, or threshold calibration.

To confirm that downstream classifiers capture genuine, transferable indicators rather than dataset artifacts, we implement four control experiments:
\begin{itemize}
    \item \textbf{Label Shuffling:} training labels are randomly permuted before feature selection and model fitting, breaking any link between speech and clinical status.
    \item \textbf{Metadata and Length Controls:} models restricted to non-speech metadata (e.g., age, gender) or trivial recording parameters (transcript length, audio duration).
    \item \textbf{Corpus Bias Control:} a baseline using only categorical corpus-identity indicators as input.
    \item \textbf{Corpus Identity Probe:} a probing classifier trained to predict source corpus from the interpretable feature space, quantifying distributional distinctiveness among the four datasets.
\end{itemize}

\section{Evidence Transferability Audit}

We first audit whether interpretable speech and language markers maintain a consistent directional relationship between HC and Risk groups across corpora and task families.

\subsection{Audit Definition}

Rather than assessing standalone significance, this audit tests the domain transferability of feature directions. For marker $m$, let $d_m^{(c)}$ denote its within-corpus standardized effect size (Cohen's $d$) between the Risk and HC groups in corpus $c$:
\begin{equation}
d_m^{(c)} = \frac{\mu_{m,1}^{(c)}-\mu_{m,0}^{(c)}}{\sigma_m^{(c)}} ,
\end{equation}
where $\mu_{m,1}^{(c)}$ and $\mu_{m,0}^{(c)}$ are the Risk and HC feature means and $\sigma_m^{(c)}$ is the pooled within-corpus standard deviation. A positive value indicates elevated marker expression in the Risk group. We define a cross-dataset direction conflict as
\begin{equation}
\gamma_m^{\mathrm{data}} = \mathbf{1}\!\left[ \min_{c\in\mathcal{C}} d_m^{(c)} < 0 \,\land\, \max_{c\in\mathcal{C}} d_m^{(c)} > 0 \right].
\end{equation}
$\gamma_m^{\mathrm{data}}$ equals 1 when a marker shows opposing directional effects across corpora, and task-family conflicts are computed analogously over task categories. A marker is deemed non-transferable if its direction reverses across deployment settings, regardless of its localized magnitude $|d_m^{(c)}|$.

\subsection{Marker Direction Conflicts}

Traditional interpretable markers rarely generalize as rigid, single-feature diagnostic rules. Among all 70 markers, 59 exhibit clear direction conflicts between HC and Risk across the four datasets, and within the 24 canonical core markers, 20 suffer cross-dataset reversals. Task-family conflicts are less frequent but still substantial: 29 of 70 markers and 7 of the 24 core markers reverse direction between picture-description tasks and connected speech.

\begin{figure}[t]
\centering
\includegraphics[width=0.97\textwidth]{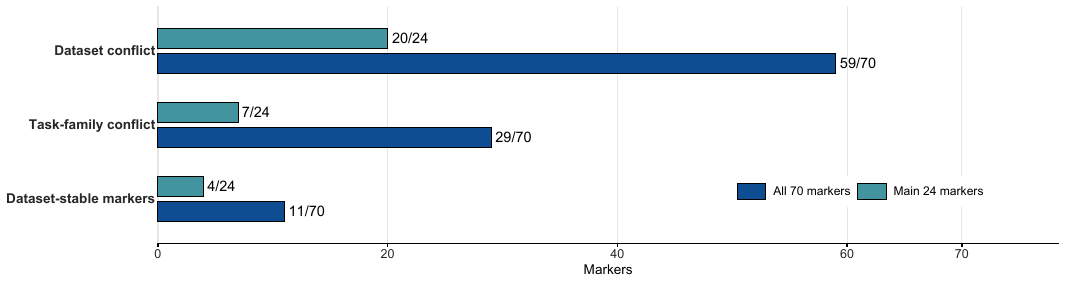}
\caption{Direction conflict prevalence across marker scopes.}
\label{fig:conflicts}
\end{figure}

Figure~\ref{fig:conflicts} quantifies these conflict distributions across marker scopes. This high conflict prevalence is a critical methodological vulnerability, since most clinical speech markers invert behavior under domain shift, which makes universal single-marker heuristics unreliable for general deployment. This directly challenges the common narrative that cognitive-decline speech is universally slower, more paused, less lexically rich, or more repetitive. While such patterns hold within specific cohorts or homogeneous tasks, our audit shows they should not be treated as invariant rules without protocol cross-verification.

Concrete features confirm this pattern. Lexical and structural markers such as pronoun ratio and function word ratio consistently elevate in Risk speech across all four corpora, while content-to-function word ratio and timestamp coverage ratio remain uniformly lower, a cross-corpus directional stability that qualifies them as reliable anchor candidates. In contrast, temporal and acoustic features such as silence ratio, mean pause duration, long pauses per minute, and character speech rate are highly sensitive to dataset and recording protocol: silence ratio decreases in the NCMMSC Risk group but increases in the other three corpora, and character speech rate elevates in the NCMMSC and Pitt Cookie Risk groups but trends downward in TAUKADIAL Mandarin and Mandarin Chou.

\subsection{Evidence Type Structure}

\begin{figure}[!ht]
\centering
\makebox[\textwidth][c]{\includegraphics[width=0.93\textwidth]{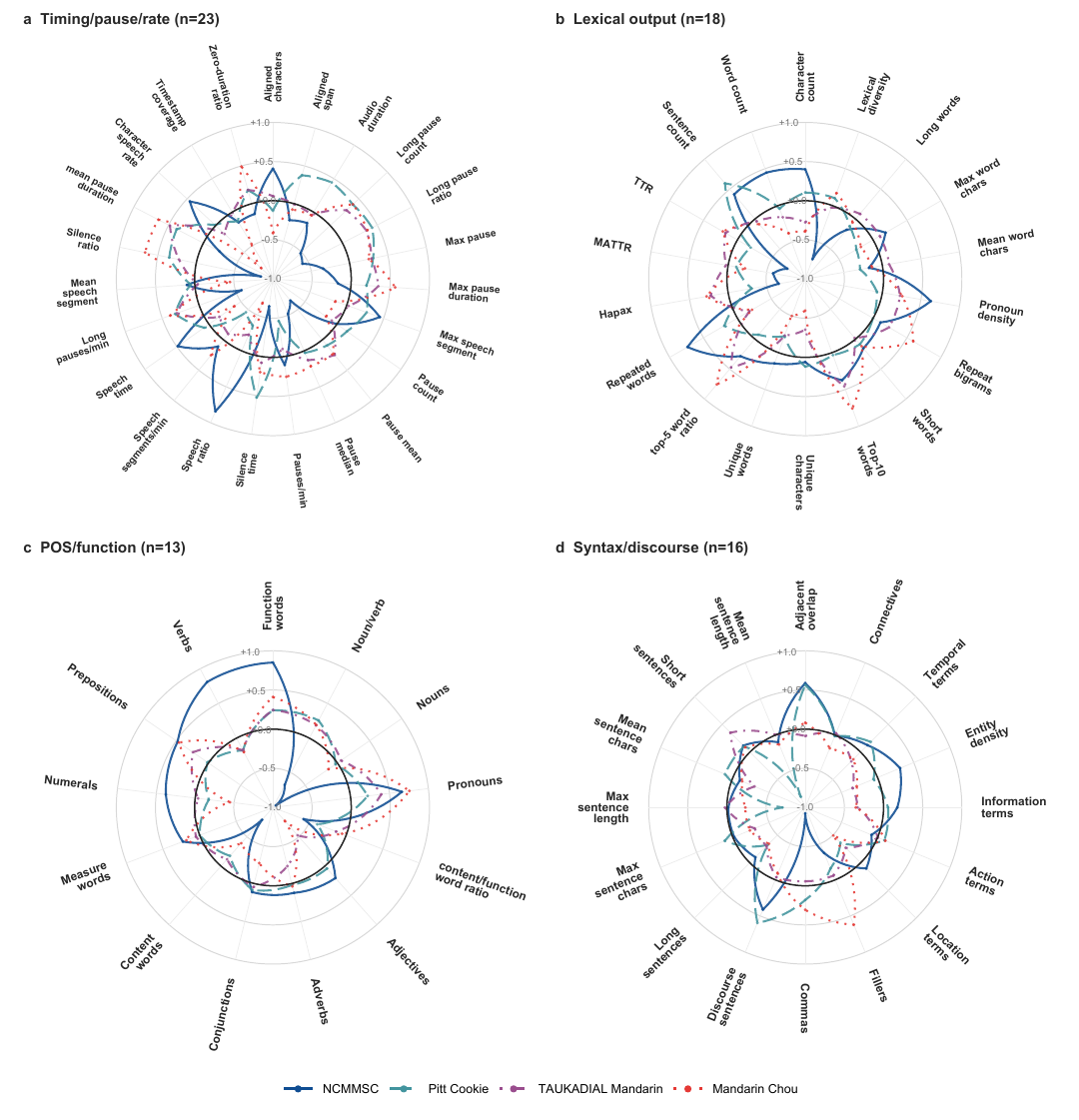}}
\caption{Grouped radar view of standardized healthy control versus risk contrasts for all 70 markers.}
\label{fig:marker_radar}
\end{figure}

The full 70-marker set is evaluated not as a classification feature space but as a baseline exposing macro-level systemic instability. Figure~\ref{fig:marker_radar} shows standardized effect sizes between HC and Risk across all four corpora, grouped by evidence type. Lexical and POS metrics align across domains, whereas temporal features such as timing, pause durations, and speech rates are highly irregular and frequently invert.

\FloatBarrier

Because features diverge directionally across domains, directly pooling all 70 markers can cause a downstream classifier to overfit to training-distribution artifacts. To mitigate this, we pre-specify a compact, interpretable panel of 24 core indicators, avoiding any feature engineering driven by held-out performance. Features are retained only if they are computable across all four corpora under a uniform speaker-level aggregation protocol, which keeps the comparison independent of corpus-specific annotation granularity. The panel is organized into nine evidence families: word class, temporal fluency, discourse structure, information content, speech quantity, lexical diversity, alignment quality, disfluency, and repetition, which separates linguistic content from acoustic timing and transcript-alignment artifacts.

\begin{figure}[!ht]
\centering
\makebox[\textwidth][c]{\includegraphics[width=0.97\textwidth]{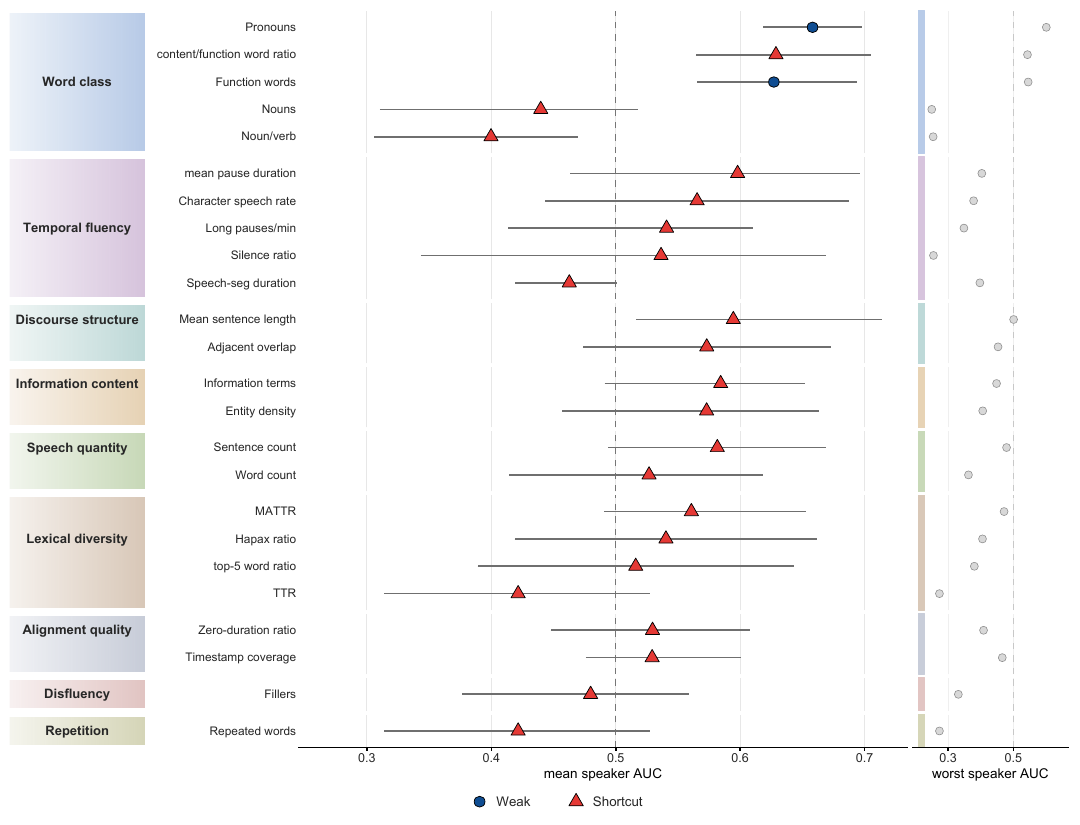}}
\caption{Single-indicator external transfer of the 24 core indicators.}
\label{fig:core24_single_marker}
\end{figure}

Figure~\ref{fig:core24_single_marker} reports the domain transferability of each core indicator evaluated in isolation under LOCO, showing the macro-averaged mean \auc\ with 95\% confidence intervals on the left and the worst-case domain \auc\ on the right. \textit{Weak} denotes a minor or inconsistent single-indicator transfer signal, while \textit{Shortcut} flags a marker that may exploit corpus-specific bias or fail catastrophically in the hardest fold. The top single indicator reaches only 0.658 mean \auc, and many features yield worst-case \auc\ below chance (0.50), reinforcing that the core 24 indicators should serve as auditable, complementary evidence channels for selection and fusion rather than standalone classifiers.

\FloatBarrier

\section{Deep Representations Under External Shift}

The marker audit raises a further question: \textbf{whether deep representations avoid these transfer vulnerabilities.} We evaluate three frozen pre-trained representations under the identical LOCO protocol, the text-based \xlmrbase\ and the acoustic wav2vec 2.0 features from the intermediate (12th) and final (24th) layers, fitting only the downstream linear classifier on the designated source corpora in each fold.

\begin{table}[!ht]
\centering
\caption{Deep-only baselines under leave one corpus out evaluation.}
\label{tab:deep_baselines}
\scriptsize
\setlength{\tabcolsep}{2.8pt}
\begin{tabular}{@{}lrrrrrrr@{}}
\toprule
Representation & NCMMSC & Pitt & TAUKADIAL & Chou & Mean & Worst & Range \\
\midrule
\xlmrbase & 0.520 & 0.739 & 0.850 & 0.968 & 0.769 & 0.520 & 0.447 \\
wav2vec 2.0 L12 & 0.462 & 0.502 & 0.835 & 0.955 & 0.688 & 0.462 & 0.493 \\
wav2vec 2.0 L24 & 0.454 & 0.574 & 0.800 & 0.969 & 0.699 & 0.454 & 0.515 \\
\bottomrule
\end{tabular}
\end{table}

\FloatBarrier

As Table~\ref{tab:deep_baselines} shows, pre-trained representations capture rich contextual information but do not inherently mitigate external domain shift. \xlmrbase\ is the strongest standalone deep baseline, reaching a mean \auc\ of 0.769, yet its fold scores range from a near-chance 0.520 on NCMMSC to a near-perfect 0.968 on Mandarin Chou. This pattern is sharper for the acoustic baselines, since both wav2vec 2.0 layers are strong on Mandarin Chou and TAUKADIAL Mandarin but collapse near chance on NCMMSC and remain weak on Pitt Cookie. Moving from the 12th to the 24th layer improves Pitt Cookie but degrades NCMMSC, showing that deeper layers alone do not resolve localized failure.

A high macro-averaged metric can therefore be driven by a few easily generalizable domains while masking catastrophic collapse elsewhere. The LOCO protocol acts as a multi-domain stress test by treating each corpus as an unseen deployment environment in turn. NCMMSC is the hardest fold here, but the broader issue is a systemic lack of worst-domain robustness under domain shift. Given its strong performance among the deep-only configurations, we select \xlmrbase\ as the primary deep network in our method, while the variances in Table~\ref{tab:deep_baselines} justify integrating an audited, interpretable evidence-selection framework.

\FloatBarrier

\section{Our Proposed Transfer Protocol}

The audit above shows that domain-transferable evidence is highly sparse and corpus-dependent. Rather than treating the full feature repository as a static input matrix, we use the audited indicators to define a constrained anchor search space, dynamically extracting robust anchors from the active training partition within each LOCO fold and fusing the resulting anchor score with the \xlmrbase\ semantic representation.

\subsection{Evidence Anchor Selection}

Evidence Anchor Selection identifies a compact, reliable feature subset using exclusively the training partition of the active LOCO fold. To limit overfitting, we restrict the candidate space to the 24 core indicators, rank them by macro-averaged training-corpus \auc\ utility, and select the top $k=4$ as domain anchors, then fit a weighted $L_2$-regularized logistic regression solely on these anchors. The held-out corpus is never observed during anchor selection, regularization tuning, or threshold calibration.

Table~\ref{tab:selection} compares this against alternative selection strategies. The source-AUC anchor selector achieves a macro-averaged mean \auc\ of 0.629 and worst-case \auc\ of 0.588, which is substantially more stable than the unselected global configurations: all 24 indicators reach only 0.521 mean \auc\ (0.418 worst), and all 70 markers reach 0.538 mean \auc\ (0.395 worst). This confirms that blindly accumulating interpretable features introduces cross-domain noise, while targeted anchor selection safeguards generalization.

Anchor composition varies by fold. For the NCMMSC fold, the training corpora select mean pause duration, pronoun ratio, top-5 word ratio, and character speech rate. For the Pitt Cookie fold, they select silence ratio, pronoun ratio, mean pause duration, and top-5 word ratio. For TAUKADIAL Mandarin, they select silence ratio, pronoun ratio, character speech rate, and content-to-function word ratio. For Mandarin Chou, they select silence ratio, pronoun ratio, content-to-function word ratio, and adjacent sentence overlap. The pronoun ratio is the only indicator selected in all four folds.

\begin{table}[t]
\caption{Traditional-indicator selection under leave one corpus out evaluation. All values are \auc.}
\label{tab:selection}
\centering
\scriptsize
\setlength{\tabcolsep}{2.5pt}
\begin{tabular*}{\textwidth}{@{\extracolsep{\fill}}lcccccc@{}}
\toprule
 & \shortstack{P-value\\top four} & \shortstack{Random\\top four} & \shortstack{Evidence-ranked\\top four} & \shortstack{All 24\\core} & \shortstack{All 70\\indicators} & \shortstack{Source-AUC\\anchors} \\
\midrule
$k$ & 4 & 4 & 4 & 24 & 70 & 4 \\
Mean \auc & 0.506 & 0.542 & 0.555 & 0.521 & 0.538 & \textbf{0.629} \\
Worst \auc & 0.363 & 0.195 & 0.520 & 0.418 & 0.395 & \textbf{0.588} \\
\bottomrule
\end{tabular*}
\end{table}

\begin{figure}[t]
\centering
\includegraphics[width=0.92\textwidth]{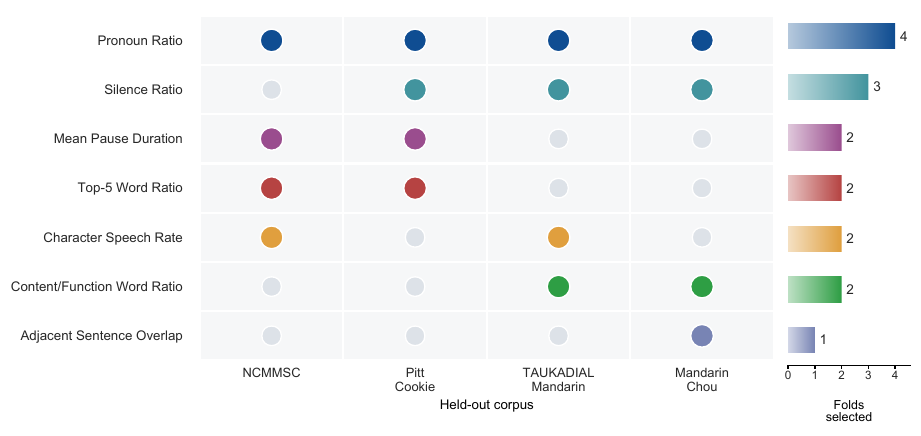}
\caption{Source-corpus anchor choices for each held-out corpus.}
\label{fig:selection}
\end{figure}

Figure~\ref{fig:selection} illustrates this cross-fold variability. The pronoun ratio serves as a persistent cross-domain invariant, while the remaining anchor channels are selected dynamically based on the source partition mixture, confirming that domain-transferable clinical evidence, though highly compact, has a composition that depends on the target deployment environment.

\subsection{Anchor Fusion}

\begin{figure}[!ht]
\centering
\includegraphics[width=0.95\textwidth]{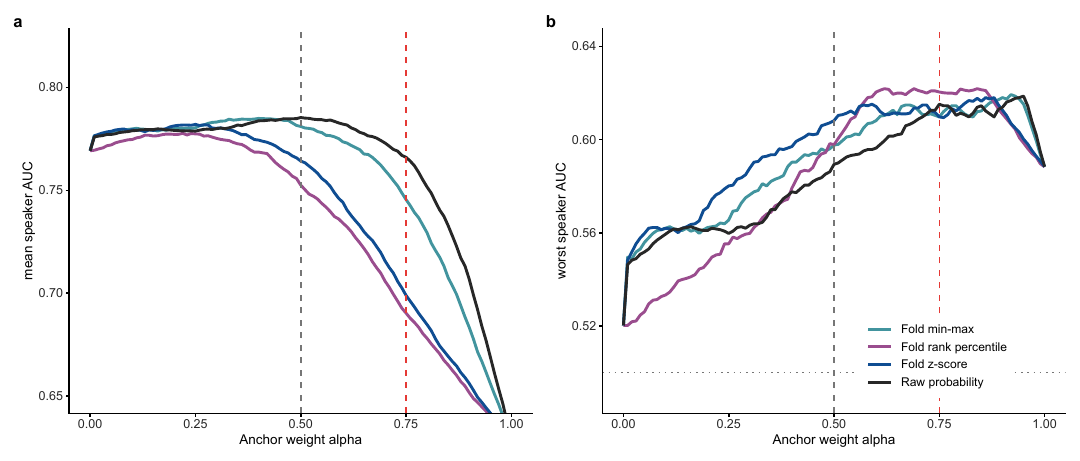}
\caption{Fusion weight sensitivity across score normalizations.}
\label{fig:alpha_sensitivity}
\end{figure}

The anchor-selection pipeline yields a compact, interpretable speaker-level score, while \xlmrbase\ captures dense, high-capacity multilingual semantic representations. We combine these under the same LOCO protocol. Let $s_{\mathrm{anchor}}$ and $s_{\mathrm{xlmr}}$ denote the speaker-level anchor and \xlmrbase\ scores, and compute
\begin{equation}
s_{\alpha}=\alpha s_{\mathrm{anchor}}+(1-\alpha)s_{\mathrm{xlmr}},
\end{equation}
where $\alpha$ is fixed before evaluating the held-out corpus and applied to speaker-averaged risk scores. We report $\alpha=0.50$ as the balanced point and $\alpha=0.75$ as the anchor-heavy point. Figure~\ref{fig:alpha_sensitivity} shows the mean-versus-weakest-corpus tradeoff across score normalizations: the balanced setting optimizes overall mean \auc, and the anchor-heavy setting favors the hardest corpus.

\begin{figure}[!ht]
\centering
\includegraphics[width=0.95\textwidth]{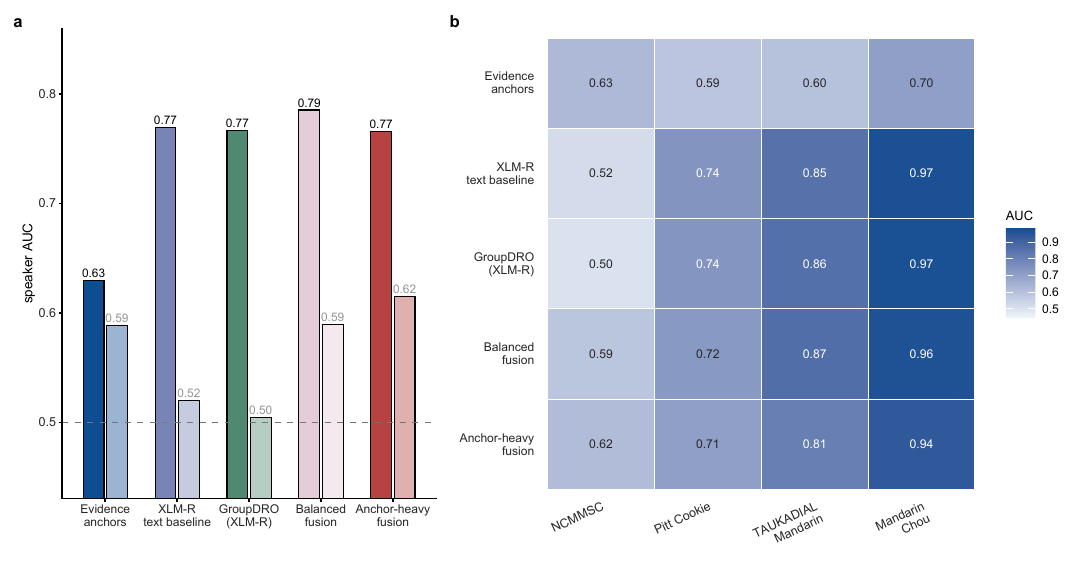}
\caption{Model comparison across held-out corpora, including the source-corpus GroupDRO domain-generalization baseline.}
\label{fig:model_comparison}
\end{figure}

Figure~\ref{fig:model_comparison} presents the primary cross-domain transfer evaluation. The anchor-only model yields lower macro-averaged performance than \xlmrbase\ alone but mitigates worst-case domain collapse, raising the lowest fold score from 0.520 to 0.588. The standard GroupDRO baseline attains a mean speaker \auc\ of 0.766 and a worst-domain \auc\ of 0.504, showing that optimizing worst source-corpus loss alone does not prevent near-chance performance on an unseen corpus. Combining both modalities, our balanced fusion ($\alpha=0.50$) achieves the top overall mean \auc\ of 0.785, 0.019 above GroupDRO, while the anchor-heavy regime ($\alpha=0.75$) secures a worst-case domain \auc\ of 0.615, 0.111 above GroupDRO and 0.095 above the deep-only baseline in the hardest deployment environment. This validates the integration of robust clinical evidence anchors.

Figure~\ref{fig:model_comparison} also shows the corpus-level \auc\ matrix behind this tradeoff. GroupDRO remains strong on Mandarin Chou (0.969), TAUKADIAL Mandarin (0.857), and Pitt Cookie (0.735), but stays near chance on NCMMSC (0.504), so robustness to observed source groups does not guarantee robustness to a new corpus. The anchors provide a more resilient floor: on the near-saturated Mandarin Chou fold, adding anchors slightly hurts performance, but on NCMMSC the anchor model is markedly more robust and lets the anchor-heavy setting substantially recover performance. TAUKADIAL Mandarin benefits most from balanced fusion, while Pitt Cookie stays close to the deep baseline.

We treat the weakest fold as an out-of-distribution stress test rather than a privileged target domain, and here NCMMSC is the bottleneck. A 2,000-resample speaker-level bootstrap gives the \xlmrbase\ an \auc\ of 0.520 (95\% CI $[0.407, 0.627]$), versus 0.615 (95\% CI $[0.509, 0.715]$) for anchor-heavy fusion. The paired bootstrap CI for this 0.095 improvement is $[-0.004, 0.195]$, confirming that heavier weight on the audited anchor stream raises the performance floor in severely shifted settings.

The two streams are also empirically distinct: the mean absolute Spearman correlation between anchor and \xlmrbase\ scores across folds is low ($\rho=0.235$, orthogonality index 0.765), and the audit phase pre-specifies which evidence channels are active in each fold before fusion with the deep stream.

Table~\ref{tab:controls} confirms these gains are not artifacts. Models trained on shuffled labels, metadata alone, or corpus-identity markers default to near-random chance, while transcript length carries negligible signal. The corpus identity probe, however, reaches high accuracy, confirming that the four datasets remain highly separable in marker space, which is why LOCO evaluation is required.

\begin{table}[t]
\caption{Negative controls and corpus separability checks under leave one corpus out evaluation. Values are mean \auc\ with 95 percent intervals.}
\label{tab:controls}
\centering
\small
\begin{tabular*}{\textwidth}{@{\extracolsep{\fill}}lcc@{}}
\toprule
Control & Mean speaker-level AUC & 95\% CI \\
\midrule
Label shuffle & 0.490 & 0.465 to 0.514 \\
Metadata only & 0.469 & 0.416 to 0.500 \\
Corpus only & 0.500 & 0.500 to 0.500 \\
Length only & 0.580 & 0.458 to 0.694 \\
Corpus identity probe & 0.899 & 0.864 to 0.934 \\
\bottomrule
\end{tabular*}
\end{table}

\FloatBarrier

\section{Discussion and Conclusion}

This study shows that real-world Alzheimer's disease speech screening is fundamentally an external domain-transfer problem, where overlapping task, linguistic, and environmental shifts cause traditional acoustic-lexical markers to reverse direction. Deep representations like \xlmrbase\ deliver high average performance but remain vulnerable to severe localized collapse on unseen domains. A source-corpus GroupDRO baseline does not close this gap under the same fixed-representation LOCO protocol. Our evidence-anchoring protocol instead gives a controllable tradeoff between mean performance ($AUC=0.785$ under balanced fusion, 0.019 above GroupDRO) and weak-domain robustness ($AUC=0.615$ under an anchor-heavy regime, 0.111 above GroupDRO). Our evaluation is limited to four English and Mandarin corpora, one generic domain-generalization baseline, a simple fusion architecture, and fixed deep representations. Even so, the results support a practical conclusion: moving speech-based cognitive screening into deployable clinical workflows requires evidence auditing and worst-case domain testing, not just stronger deep classifiers or source-loss reweighting.

\subsubsection{Disclosure of Interests.} The authors have no competing interests to declare that are relevant to the content of this article.

\bibliographystyle{splncs04}
\bibliography{references}

\end{document}